\documentclass[prb,twocolumn,floatfix,showpacs,preprintnumbers,superscriptaddress,amsmath,amssymb]{revtex4-1}

\usepackage{graphicx}
\usepackage{dcolumn}

\begin{document}

\title{Field enhancement in subnanometer metallic gaps}

\author{A. Garc\'{\i}a-Mart\'{\i}n}
\affiliation{IMM-Instituto de Microelectr\'onica de Madrid (CNM-CSIC),
Isaac Newton 8, PTM, Tres Cantos, E-28760 Madrid, Spain.}

\author{D. R. Ward}
\affiliation{Department of Physics and Astronomy, Rice University, 
6100 Main St., Houston, TX 77005 USA}

\author{D. Natelson}
\affiliation{Department of Physics and Astronomy, Rice University, 
6100 Main St., Houston, TX 77005 USA}
\affiliation{Department of Electrical and Computer Engineering, 
Rice University, 6100 Main St., Houston, TX 77005 USA}

\author{J. C. Cuevas}
\affiliation{Departamento de F\'{\i}sica Te\'orica de la Materia Condensada,
Universidad Aut\'onoma de Madrid, E-28049 Madrid, Spain.}

\date{\today}

\begin{abstract}
Motivated by recent experiments [Ward \emph{et al.}, Nature
Nanotech. {\bf 5}, 732 (2010)], we present here a theoretical
analysis of the optical response of sharp gold electrodes
separated by a subnanometer gap. In particular, we have used
classical finite difference time domain simulations to
investigate the electric field distribution in these nanojunctions
upon illumination.  Our results show a strong confinement of the
field within  the gap region, resulting in a large enhancement
compared to the incident field. Enhancement factors exceeding
$10^3$ are found for interelectrode distances on the order of a
few \AA, which are fully compatible with the experimental
findings.  Such huge enhancements originate from the coupling of
the incident light to the evanescent field of hybrid plasmons
involving charge density oscillations in both electrodes.
\end{abstract}


\maketitle

\emph{Introduction.--} The study of the optical response of
metallic nanostructures is revealing fascinating new 
physics.\cite{Maier2007,Schuller2010}  Special attention is being
paid to the analysis of the so-called optical and infrared 
\emph{gap nanoantennas}, which consist of adjacent metallic segments,
like nanorods, separated by a nanoscale gap.\cite{Grober1997,
Muehlschlegel2005,Schuck2005,Tang2008,Kim2008,Ghenuche2008,
Schnell2009,Schnell2010,Bharadwaj2009} The ability of these
systems to efficiently confine and enhance optical
fields is crucial for applications such as single-molecule
surface-enhanced Raman spectroscopy (SERS) \cite{Xu1999} or
extreme-ultraviolet generation.\cite{Kim2008}

There are many techniques to measure the
near-field distributions in gap nanoantennas such as scanning
near-field optical microscopy \cite{Schnell2009,Schnell2010} and
two-photon induced luminescence.\cite{Muehlschlegel2005,Ghenuche2008} 
However, these local probes have a limited spatial resolution of 10 
nm at best.\cite{Schnell2010} It would be thus highly desirable to 
develop new techniques or strategies that enable the extraction of 
information about the local fields in subnanometer metallic gaps, 
where the field enhancements are expected to be largest. Very recently,
we made a step in this direction,\cite{Ward2010} investigating the 
electronic transport through atomic-scale gold electrodes separated 
by a subnanometer gap under near-infrared laser irradiation, finding
that the irradiation induces a d.c.\ photocurrent. By comparing
this photocurrent with low-frequency conduction measurements, we
were able to determine the optical voltage generated across the
gap and, in turn, to infer the electric field in this region.
Enhancement factors exceeding $10^3$ were reported, in
line with previous estimates from surface-enhanced Raman
measurements.\cite{Ward2007,Ward2008a,Ward2008b} The goal of this
Communication is to shed some light on the origin of this huge
field enhancement in these atomic-scale gap antennas.

From a theoretical point of view, the field enhancement in finite
metallic structures with nanometric gaps such as nanoparticle or
nanorod dimers has been extensively studied (see
Ref.~\onlinecite{Pelton2008} and references therein). However,
here we are interested in the analysis of a sub-nanometer gap
formed between two atomic-scale electrodes, thus coupled to
semi-infinite leads, where, to our knowledge, no related studies
have been reported.\cite{note1,note2} In this work we present an 
analysis of the optical response of gold atomic junctions with 
subnanometer gaps based on finite difference time domain (FDTD) 
simulations. Our main findings are: (i) field enhancements exceeding 
$10^3$ are possible for gaps of a few \AA,\cite{Li2003}
which supports the main conclusion of Ref.~\onlinecite{Ward2010}; 
(ii) the huge enhancements are due to the excitation of a hybrid 
plasmon involving large localized charge distributions of opposite 
sign on either side of the junction, in analogy with the dipolar 
bonding dimer plasmons found in nanoparticle 
dimers;\cite{Romero2006,Perez-Gonzalez2010} and (iii) the plasmon 
resonances red-shift as the interelectrode gap decreases, which 
also resembles the behavior found in nanoparticle \cite{Romero2006} 
and nanorod dimers.\cite{Aizpurua2005}

\begin{figure*}[t]
\begin{center}
\includegraphics*[width=\textwidth,clip]{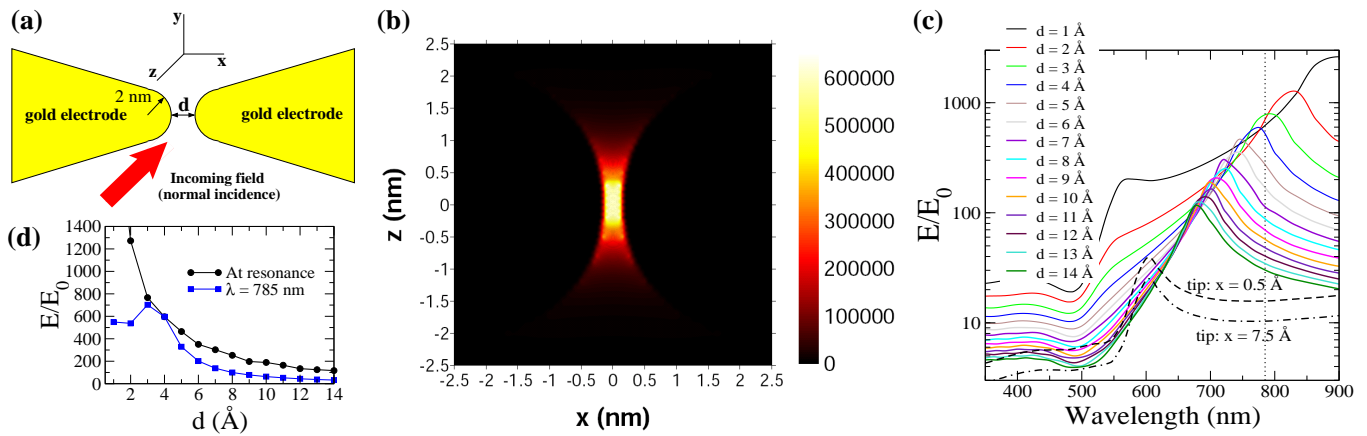}
\caption{\label{spectral} (Color online) (a) Schematic
representation of the gold junction considered in this work. (b)
Spatial distribution of the field intensity (normalized by the
incident one) in the $x$-$z$ plane for an interelectrode
separation $d=3$ \AA.  The wavelength of the illuminating light is
$\lambda=801$ nm. (c) Field enhancement ($E/E_0$) evaluated in the
middle of the gap as a function of the wavelength of the incident 
light for different interelectrode separations. In all cases the
polarization is directed along the $x$ axis (junction axis). The
dashed and dashed-dotted lines are the corresponding results for a
single tip at distances 0.5 \AA \, and 7.5 \AA \, from the tip
along the $x$ axis, respectively. The vertical dotted line
indicates the wavelength 785 nm used in Ref.~\onlinecite{Ward2010}. 
(d) Field enhancement in the middle of the gap as a function of the 
interelectrode distance at resonance and at $\lambda=785$ nm.}
\end{center}
\end{figure*}

\emph{System and methodology.--} We model here the gold nanogaps
of Ref.~\onlinecite{Ward2010} with the idealized geometry of
Fig.~\ref{spectral}(a). In this geometry, two extended gold tips ending 
in semispheres with a radius of 2 nm are separated by a distance $d$
(this distance is assumed to be the actual distance between the
electron clouds).  We also consider that the junction is placed in
vacuum and that there is no substrate.  To analyze the optical
response of this gold junction we have performed FDTD simulations
where the main features to be considered are: (i) the two tip-like
electrodes are coupled to infinite metallic surfaces, which are 
described by perfect metallic boundary conditions;\cite{note3} (ii) 
the structure is illuminated at normal incidence by a plane wave
covering the whole region, and perfectly absorptive boundary
conditions are placed on the illumination direction; (iii) the
gold dielectric function used was extracted from ellipsometry
measurements of a 20~nm thin film;\cite{Elias2009} and (iv) a
non-uniform mesh with a smallest grid size of 0.25~\AA \ at the 
junction was used. All the simulations were performed with the
code FDTD Solutions (from Lumerical Solutions, Inc., Canada). 

\emph{Results and discussions.--} In Fig.~\ref{spectral}(b) we show
the spatial distribution of the electric field intensity (normalized
by the incident intensity) in the $x$-$z$ plane of a junction with
$d=3$~\AA \ illuminated by an incident light of 801~nm with its
polarization along the junction axis ($x$ axis).  The field is
strongly localized in the gap region, and the intensity (i.e.\ $|E|^2$)
is enhanced by a factor larger than $6 \times 10^5$ in that region.
We have systematically analyzed the field distribution as a function
of both the interelectrode distance $d$ and the wavelength of the
incident light. A summary of the results can be seen in
Fig.~\ref{spectral}(c), where we show the field enhancement factor
($E/E_0$) in the middle of the gap on the junction axis, with $E_0$
the amplitude of the incident field, as a function of the wavelength,
and for different values of $d$ ranging from 1 to 14~\AA.  Notice the
appearance of a resonance that shifts monotonically to the red as
$d$ decreases.  At large separations, this resonance wavelength tends
toward that exhibited by a \textit{single} tip, around 600 nm (see
dashed and dashed-dotted lines). On the other hand, the enhancement
factor on resonance reaches values larger than $10^3$ for $d <3$~\AA,
which supports the estimates reported in Ref.~\onlinecite{Ward2010}.

The resonance seen in Fig.~\ref{spectral}(c) clearly suggests that
the incident light is exciting a plasmon-type mode. The
comparison of the wavelength of that resonance with the one of a
single tip and the fact the resonance red-shifts as $d$ decreases
both indicate that this mode can be considered as a hybrid plasmon
involving charge density oscillations of opposite signs on both
sides of the junction. The red shift is a simple consequence of
the increasing interelectrode interaction as $d$ decreases, which
in turn leads to a reinforcement of the electric field in the gap
region. It is worth stressing that in this case the hybridization 
occurs between the continua of delocalized plasmon modes of the 
extended tips, rather than between localized plasmons as in the 
case of subwavelength nanoparticles (see discussion below).

The actual value of the field enhancement factor depends on the 
tip radius and it increases monotonically as the radius decreases.
This is precisely the well-known ``lightning-rod'' effect, see
\textit{e.g.}\ Ref.~\onlinecite{Gersten1980}.  In this case the
lightning-rod effect cooperates with the plasmon excitation to 
greatly enhance the field locally at the region. To get an idea of 
the impact of this effect, we have repeated the calculations with 
a smaller tip radius of 1.5 \AA \, and found that, while the 
spectral response is almost identical, the field maximum increases 
by approximately a 16\%.

In Ref.~\onlinecite{Ward2010} it was found that the enhancement
decreases slowly with $d$ (slower than $1/d$).  Our simulations
show that this decay depends critically on the wavelength of the
incident light. In Fig.~\ref{spectral}(d) we show the enhancement
factor in the middle of the gap as a function of $d$ both for the
resonant wavelength and for 785 nm, the wavelength used
in Ref.~\onlinecite{Ward2010}. Notice that for the resonant
condition, the field decays monotonically with the gap size
approximately as $1/d^{1.24}$, while for $\lambda = 785$ nm the
field does not decay for very short distances, and for $d > 4$ \AA
\, decays slightly faster than $1/d^{2}$.

\begin{figure}[t]
\begin{center}
\includegraphics*[width=\columnwidth,clip]{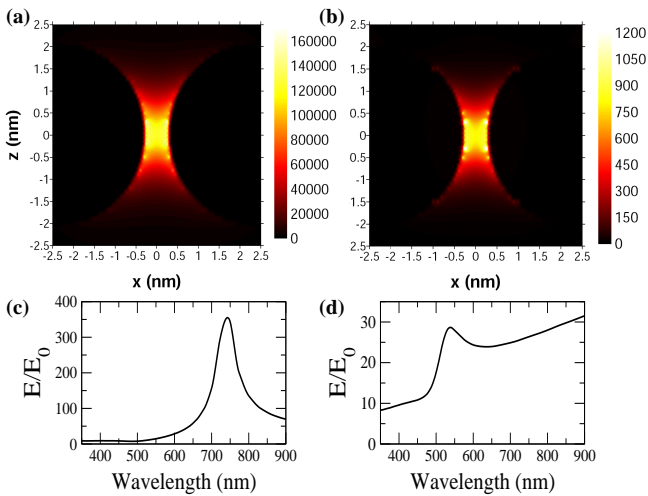}
\caption{\label{spheres} (Color online) (a) Spatial distribution of the
normalized field intensity in the gold junction for $d=6$ \AA \ and
$\lambda = 750$ nm. (b) The same as in panel (a), but for two spheres
of radius 2 nm separated 6 \AA \ and illuminated with light of 539 nm.
(c) Field enhancement ($E/E_0$) evaluated in the middle of the gap of the
gold junction of panel (a) as a function of the light wavelength. (d) The
same as in panel (c), but for the nanoparticle dimer of panel (b).}
\end{center}
\end{figure}

The results described above qualitatively resemble the predicted and
observed optical response of nanoparticle dimers (see e.g.\
Ref.~\onlinecite{Romero2006}) and nanorods dimers (see e.g.\
Ref.~\onlinecite{Aizpurua2005}).  For instance, it is well-known that
when two nanoparticles are placed next to each other, the plasmon
modes of the individual nanoparticles interact, resulting in
hybridized dimer plasmon modes whose energies can be strongly
red-shifted with respect to that of the plasmon modes of the
individual nanoparticles.  This is essentially what happens in the
gold junctions considered here.  One may then wonder to what extent
the field enhancement in gold junctions can be explained in terms of
the physics of nanoparticle dimers. To answer this question, we have
considered a dimer comprising two gold spheres of radius 2 nm, which
is the radius of the outermost part of the gold tips.\cite{note4} In the 
upper panels of Fig.~\ref{spheres}, we compare the field distributions for 
a gold junction with $d=6$ \AA \ (panel (a)) and a dimer with the same
separation (panel (b)). In each case the wavelength considered is the
one that gives rise to the maximum field (750 nm for the junction and
539 nm for the dimer). Notice that while the symmetry of the field
distribution in the gap region is practically identical in both cases,
there is a big difference in the magnitude of the field enhancements
(of more than an order of magnitude). Moreover, as we show in 
Fig.~\ref{spheres}(c-d), there is also an important difference in the
spectral response, where the resonance for the extended electrode junction
is considerably red-shifted as compared with the dimer case. The smaller 
field enhancement for the dimer is mainly due to its smaller scattering 
cross-section. Of course, in the dimer configuration one could reach 
the enhancement factors of the junction for the same gap size, but 
that would require the nanoparticles to have a considerably larger
radius,\cite{Aubry2010} which would be unrealistic for an atomic-size 
contact. Although less important, the difference in the field values
reached at the resonant conditions in both structures is also partly due 
to the frequency dependence of the gold permittivity, which in particular
has a slightly larger imaginary part at the frequency of the plasmon 
resonance of the dimer.\cite{note5}

\begin{figure}[t]
\begin{center}
\includegraphics*[width=0.7\columnwidth,clip]{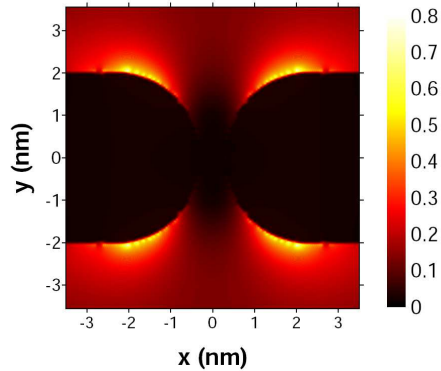}
\caption{\label{spol} (Color online) (a) Spatial distribution of the
normalized field intensity in the $x$-$y$ plane for a gold junction 
with $d=6$ \AA \ and illuminated with light of 495 nm with its polarization 
directed along the axis perpendicular to the junction axis ($y$-axis).
The wavelength used is the one at which the field reaches its maximum in 
the middle of the junction.}
\end{center}
\end{figure}

The plasmonic origin of the field enhancement suggests that it should
be very sensitive to the polarization of the incident field.\cite{Tian2006}
This is indeed the case, as we illustrate in Fig.~\ref{spol}. This 
figure shows the intensity distribution in a junction with electrode
separation $d=6$ \AA \ and $\lambda = 495$ nm, but this time the
polarization is directed along the axis perpendicular to the junction
axis (this figure has to be compared with Fig.~\ref{spheres}(a)).  As
one can see, the field distribution is now quite different (with the
charge oscillating back and forth in the transversal direction) and,
in particular, the near-field is smaller than the incident field
everywhere in the junction region. Notice that a field distribution
like this does not generate an optical voltage across the junction and
therefore, there would not be any photocurrent in a transport experiment 
like the one of Ref.~\onlinecite{Ward2010}. We have also verified that 
the strong polarization dependence of the field enhancement persists 
even in very asymmetric contacts with a pronounced misalignment of 
the axes of the tips. Further investigation into other tip geometries 
are ongoing.

It is important to discuss the limitations of our classical theory.
First, we have assumed that the optical response of these nanocontacts
is well described by a classical frequency-dependent dielectric
function. However, non-local effects can play some role in the
outermost part of the electrodes, when the dimensions are smaller than
the mean free path of the valence electrons that participate in the
plasmon resonances.\cite{Fuchs1990} Second, the onset of quantum
effects when the tips are very close can be much more important. As
discussed in Ref.~\onlinecite{Zuloaga2009}, the onset of quantum
tunneling and the concomitant charge transfer between the electrodes
can lead to the appearance of a charge transfer plasmon involving
conduction electrons flowing back and forth between the metallic tips.
A finite electron density between the tips could give rise to a
screening of the plasmonic interactions responsible for the strong red
shift of the dipolar plasmons and, in turn, could reduce the field in
the gap region.\cite{Mao2009} In any case, while our classical approach 
neglects detailed electronic structure, the essential physics of plasmon 
hybridization involving the continuum modes of the extended electrodes 
is expected to survive intact in realistic quantum mechanical calculations.

\emph{Conclusions.--} We have studied the optical response of
atomic-scale gold junctions with subnanometer gaps, within the
framework of classical electromagnetism.  We have shown that the huge
field enhancements reported experimentally \cite{Ward2010} originate
from the excitation of hybrid plasmons resonances involving charge
oscillations in both electrodes.  Such resonances red-shift as the gap
size decreases as a consequence of the increase of the interelectrode
interaction.  Our results on the magnitude of the field enhancement
clearly indicate that metallic nanogaps can be ideal templates for
surface-enhanced Raman scattering of single molecules, very important
for both molecular electronics and sensing
applications.\cite{Cuevas2010}

\emph{Acknowledgments.--} We thank F.J. Garcia-Vidal, J.J. S\'aenz and
J.K. Viljas for valuable discussions. A.G.M.\ acknowledges financial support 
from the EU (NMP3-SL-2008-214107-Nanomagma), and the Spanish MICINN (``MAGPLAS"
Grant No. MAT2008-06765-C02-01/NAN and ``FUNCOAT" CONSOLIDER INGENIO
Grant No. 2010 CSD2008-00023). J.C.C.\ acknowledges support from the
EU through the BIMORE network (grant MRTN-CT-2006-035859) and the
Spanish MICINN (grant FIS2008-04209). D.N. and D.R.W. acknowledge
support from the Robert A. Welch Foundation (grant C-1636) and the
Lockheed Martin Advanced Nanotechnology Center of Excellence at Rice
(LANCER).



\begin{thebibliography}{10}
\bibitem{note3}
The infinite surfaces are placed 8 nm away from the center of the junction.
\bibitem{note4}
For the description of the nanoparticle dimer we replaced the infinite
metallic surfaces used for the junction by perfectly absorptive boundary 
conditions.
\bibitem{note5}
The gold dielectric constant in our case is $\epsilon = -6.28 + 2.04 i$
for $\lambda = 539$ nm and $\epsilon = -23.46 + 1.63 i$ for $\lambda = 750$
nm.
\end{thebibliography}
\end{document}